\documentclass[12pt]{article}
\usepackage{graphicx}

\def\be{\begin{equation}}
\def\ee{\end{equation}}
\def\ba{\begin{eqnarray}}
\def\ea{\end{eqnarray}}

\def\ii{{\rm i}}
\def\tr{{\rm tr}}

\def\diag{{\rm diag}}

\newcommand{\eq}{\begin{equation}}
\newcommand{\en}{\end{equation}}
\newcommand{\eqa}{\begin{eqnarray}}
\newcommand{\ena}{\end{eqnarray}}

\newcommand{\NP}[1]{Nucl.\ Phys.\ {\bf #1}}

\newcommand{\MPL}[1]{Mod.\ Phys.\ Lett.\ {\bf #1}}
\newcommand{\IJMP}[1]{Int.\ J.\ Mod.\ Phys.\ {\bf #1}}

\def\hepth#1{{\tt hep-th/#1}}
\begin{document}
\begin{titlepage}
\vskip0.5cm
\begin{flushright}
DFTT 63/99\\
\end{flushright}
\vskip0.5cm
\begin{center}
{\Large\bf
Matrix strings from generalized Yang-Mills theory on arbitrary
Riemann surfaces.
}
\end{center}
\vskip 1.cm
\centerline{
M. Bill\'o$^a$,
A. D'Adda$^a$ and P. Provero$^{b,a}$}
\vskip0.6cm
\centerline{\sl  $^a$ Dipartimento di Fisica
Teorica dell'Universit\`a di Torino and}
\centerline{\sl Istituto Nazionale di Fisica Nucleare, Sezione di Torino}
\centerline{\sl via P.Giuria 1, I-10125 Torino, Italy}
\vskip 0.2cm
\centerline{\sl $^{b}$ Dipartimento di Scienze e Tecnologie Avanzate}
\centerline{\sl Universit\`a del Piemonte Orientale}
\centerline{\sl I-15100 Alessandria, Italy
\footnote{e--mail: {\tt billo, dadda, provero@to.infn.it}}}
\vskip 0.6cm
\begin{abstract}
We quantize pure 2d Yang-Mills theory on an arbitrary Riemann surface in the
gauge where the field strength is diagonal. Twisted sectors originate, as
in Matrix string theory, from permutations  of the eigenvalues around
homotopically non-trivial loops.
These sectors, that must be discarded in the usual quantization due to
divergences occurring when two eigenvalues coincide, can be
consistently kept if one modifies the action by introducing a coupling of
the field strength to the space-time curvature.
This leads to a generalized Yang-Mills theory whose action reduces to the
usual one in the limit of zero curvature.
After integrating over the non-diagonal components of the gauge fields, the
theory becomes a free string theory (sum over unbranched coverings) with a
U$(1)$ gauge theory on the world-sheet. This is shown to be equivalent to
a lattice theory with a gauge group which is the semi-direct product of
$S_N$ and U$(1)^N$. By using well known results on the statistics
of coverings, the partition function on arbitrary Riemann surfaces and
the kernel functions on surfaces with boundaries are calculated.
Extensions to include branch points and non-abelian groups on
the world-sheet are briefly commented upon.
\end{abstract}
\end{titlepage}
\setcounter{footnote}{0}
\def\thefootnote{\arabic{footnote}}
\section{Introduction}
The solution of pure two-dimensional Yang-Mills theory (YM2 in the
following) on a general Riemann surface has been known for some time.
It was obtained on the lattice in \cite{m75,r90}, and in the continuum
in \cite{w91,wit92,bt92}. For instance, the partition
function of the U$(N)$ theory on a surface of genus $G$ and area ${\cal A}$
is
\begin{equation}
\label{intro1}
Z = \sum_R (d_R)^{2-2G}\, {\rm e}^{-{1\over 2}g^2{\cal A}\, C_R}~,
\end{equation}
where the sum runs on the irreducible representations $R$, of dimension $d_R$
and quadratic casimir $C_R$,  of the U$(N)$ group.
\par
This solution allowed  Gross and Taylor \cite{gt} to describe the
U$(N)$ YM2 theory, in the large $N$ limit, as a particular type of string
theory. In fact they could show that in this limit
the partition function (\ref{intro1}) counts distinct
maps from world-sheets of genera $g$ to the target space represented by the
base manifold of genus $G$.
\par
Recently, a different way to formulate  YM2 on a torus or a cylinder
as a string theory has been suggested \cite{us,uscorfu}, and in this paper we
are going to extend this formulation to arbitrary Riemann surfaces.
This proposal is inspired by Matrix string theory \cite{motl,dvv}.
In Matrix string theory one considers the ${\cal N}=8$ supersymmetric
YM2 theory, defined on a cylinder. This theory contains eight
scalar fields $X_M$ with their supersymmetric partners $\Theta_{\alpha}$
in the adjoint representation of the U$(N)$ gauge group.
In the limit
$g_{\rm YM}\to\infty$ all these fields commute and can be
simultaneously diagonalized. The $N$ eigenvalues of $X_M$
take the role of the eight transverse coordinates describing the
world-sheets of a gas of free superstrings. The eigenvalues, however, can be
subjected to a permutation as one goes around the compact direction of the
cylinder,
giving rise to twisted sectors in correspondence with the conjugacy classes of
the symmetric group $S_N$.
Strings of different lengths are present in each sector, and correspond to the
cycles of the permutation.
\par
In~\cite{us} we showed that twisted sectors appear naturally in pure YM2 on a
torus or a cylinder if one chooses the so-called {\em unitary gauge}~\cite{blth}, namely the gauge in which the field strength is diagonal. Their
inclusion leads to a partition function that is different from the standard one
of~\cite{r90}, has a richer structure and is closely related to the partition
function of the whole matrix string theory~\cite{kv,Grignani:1999sp}.
\par
However, in higher genus Riemann surfaces there are always divergences 
associated to the twisted sectors in correspondence of points where two 
eigenvalues of the field strength coincide. 
A rather {\it ad hoc} regularization was suggested
in \cite{bt92}, whose effect is to suppress all non-trivial sectors in order to
reproduce the standard result.
\par
We propose here to study a generalized YM2
(in the sense of \cite{dls94,gsy95})
in which such
divergences do not occur, and the twisted sectors can be consistently
kept into account. This can be achieved by introducing a suitable
coupling of the field-strength to the Riemann curvature of the space-time
manifold\footnote{This coupling can be interpreted as the result of the
gauge fields being defined on the tangent space (like fermions) rather than
being one-forms on the base manifold}. The action of the generalized theory
reduces to the one of ordinary YM2 for flat surfaces.
In the generalized theory the twisted sectors are in one-to one
correspondence with the unbranched $N$-coverings of the surface, and the
original U$(N)$ gauge group reduces to an abelian gauge group for each
connected component
of the covering. For the torus and the cylinder the results of~\cite{us}
are reproduced. All previous results are discussed in Sections 2 and 3.
In Section 4 we show that, after integration over the non Cartan components of
the gauge fields, the theory can be  described as a lattice gauge theory of a
group ${\cal G}_N$ defined as the semi-direct product of $S_N$ and U$(1)^N$.
By using known results on the statistics of coverings~\cite{gt,ksw}, we derive
in Sections 5 and 6 the expression for the partition function on surfaces
without boundaries and the kernel
function for surfaces with an arbitrary number
of boundaries.
In Section 7, with the conclusions, we outline the main features of the
extension of the present theory to the case of interacting strings (branched
coverings) and to strings with a non-abelian gauge group on the world-sheet.
\section{Unitary gauge and Matrix string theory}
Consider a two-dimensional gauge theory defined on a
Riemann surface ${\cal M}$,
with or without boundaries,
and with gauge group  ${\cal G}$, for instance U$(N)$. Assume that the theory
contains at least one field $F$ transforming in the adjoint
representation of ${\cal G}$; in the case of U$(N)$, $F$ is a $N\times N$
hermitian matrix.
A possible gauge choice consists in choosing $F$ with values in the
Cartan subalgebra of  ${\cal G}$, that is, in the U$(N)$ case, in choosing $F$
diagonal (unitary gauge). In fact it is always possible
at each point $x$ of ${\cal M}$ to find a gauge
transformation $g(x)$ that casts $F$ into a diagonal form:
\begin{equation}
g^{-1}(x) F(x) g(x) = \diag(\lambda(x))~.\label{Pgaugefix}
\end{equation}
Such gauge fixing is not complete as it leaves a U$(1)^N$ gauge
invariance, and is affected by Gribov ambiguities because
at each point $x$ there are in general $N!$ different gauge-fixed forms of
$F$, related to each other by arbitrary permutations of the eigenvalues:
\begin{equation}
Q^{-1} \lambda_i (x) Q =\lambda_{Q(i)}(x)~.
\label{permutation}
\end{equation}
\par
As a consequence of the Gribov ambiguity it is not possible in general
to put $F$ in diagonal form globally, namely on the whole ${\cal M}$
without discontinuities. In fact if we map ${\cal M}$  into a polygon
with pairs of edges suitably identified, and  fix the unitary gauge on the
polygon (this is always possible as the bundle is trivial on a disc),
after gauge fixing the eigenvalues of $F$ will in general appear in a
different order on the edges that are to be identified.
\par
More precisely: consider a point $x$ on ${\cal M}$, and let $\Pi_1 (x|{\cal
M})$  be the homotopy group on ${\cal M}$ with base point $x$. Let us order
the eigenvalues of $F$ in $x$ in an arbitrary way, for instance
\begin{equation}
\label{standardorder}
\lambda_1(x)>\lambda_2(x)>....>\lambda_N(x)~.
\end{equation}
This will be referred to in the rest of the paper as the {\it standard order}.
Consider now  a non trivial element $\gamma_x$ of $\Pi_1 (x|{\cal M})$, that
is a homotopically non-trivial loop starting from $x$.
As we go round the loop the eigenvalues of $F$ are subjected to a permutation,
or, for a generic gauge group, they are acted upon by an element of the Weyl
group.
This defines a homomorphism that associates to each element $\gamma_x$ of the
homotopy group $\Pi_1 (x|{\cal M})$  an element $P(\gamma_x)$ of the Weyl
group.
\par
The permutation $P(\gamma_x)$ depends on the choice of the base point $x$
according to the formula
\begin{equation}
P(\bar{\gamma}_y) = P(\gamma^{-1}_{y \rightarrow x}) P(\gamma_x) P(\gamma_{y
\rightarrow x})~,
\label{conj}
\end{equation}
where $\bar{\gamma}_y $ is a closed loop starting in $y$ and defined by
$\bar{\gamma}_y= \gamma^{-1}_{y \rightarrow x} \gamma_x
\gamma_{y \rightarrow x}$ where $ \gamma_{y \rightarrow x}$ is an arbitrarily
chosen path from $y$ to $x$.  $P(\gamma_{y \rightarrow x})$ is the permutation
of the eigenvalues in $x$ with respect to the standard order defined above
obtained by starting with the eigenvalues in the standard order in
$y$ and moving to $x$ along the chosen path $ \gamma_{y \rightarrow x}$.
Eq. (\ref{conj}) shows that changing the base point of the homotopy group,
namely deforming the original path $\gamma_x$ into $\bar{\gamma}_y $,
corresponds to a conjugacy transformation on $ P(\gamma_x)$.
\par
According to this analysis the functional integral over the field
configurations of $F$ splits into several topological sectors, one for
each homomorphism of $\Pi_1 (x|{\cal M})$ into $S_N$.
\par
It is easy to see that in the case of U$(N)$, where the eigenvalues of $F$ are
real and the Weyl group is the symmetric group $S_N$, the homomorphic maps
of $\Pi_1 (x|{\cal M})$ into $S_N$ are in one-to-one correspondence with
the $N$-coverings of ${\cal M}$ without foldings or branch points.
In this correspondence each eigenvalue of $F$ is associated to
one sheet of the $N$-covering.
Such coverings are not in general connected but consist of a certain number of
connected parts. The decomposition of a covering into its connected parts
coincides with the decomposition of the representations of $\Pi_1 (x|{\cal M})$
into its irreducible subspaces.
A connected part that covers ${\cal M}$ $k$
times consists of a $k$-dimensional subspace of the $N$-dimensional space
of the eigenvalues of $F$. This subspace is invariant under the action of
$\Pi_1 (x|{\cal M})$  and does not contain any lower dimensional invariant
subspace.
\par
Each connected part can be thought of as the world-sheet of a two-dimensional
string that covers $k$ times the target space and that can be described as
follows.
Let us denote by $\xi_{\alpha}$ ($\alpha=1,2$) a set of coordinates
parameterizing the world-sheet. Each  point $x$ of the target space has $k$
images on the world-sheet labeled by $\xi_{\alpha,i}(x)$ with $i=1,2,\dots,k$.
So we can associate to each point of the world-sheet a single eigenvalue
$\lambda(\xi)$ by requiring
\begin{equation}
\lambda(\xi_{i}(x)) = \lambda_i(x)~.
\label{worldsheet}
\end{equation}
In the end we have a single eigenvalue and a U$(1)$ gauge invariance associated
to each connected part of the covering, namely to each irreducible part of the
representation of $\Pi_1 (x|{\cal M})$ in terms of elements of $S_N$.
\par
We may be tempted to conclude that, through the mechanism outlined above, a non
abelian gauge theory in two dimensions can be described in terms of a string
theory with an abelian gauge theory on its world-sheet.
However this is true only if the abelianization is not spoiled by other terms
in the action that induce a coupling between different sheets of the covering.
Therefore we have to require that, at least in some limit for the coupling
constants, the strings described above are free or weakly coupled.
\par
In this respect a problem arises directly from the gauge fixing procedure.
Indeed, the Faddeev-Popov determinant for our gauge choice induces
a strong interaction between different sheets of the covering. In fact,
the functional integration measure produces, as usual,  a squared Vandermonde
determinant of the eigenvalues of $F$:
\begin{equation}
{\cal D}F(x) =  {\cal D}\lambda(x) \prod_{x} \Delta (\lambda)^2 =
{\cal D}\lambda(x) \exp \left\{ 2 \int d\mu \sum_{i>j} \log 
|\lambda_i(x) - \lambda_j(x) |  \right\}~,
\label{fadpop}
\end{equation}
where $d\mu = d^2x \sqrt{g}$ is the area element.
This term is singular when two eigenvalues coincide and indeed it describes an
interaction of different world-sheets ({\em i.e.} eigenvalues) at the same 
target space point $x$.
Therefore we must require that the Vandermonde determinants
arising from the gauge fixing are exactly canceled for the string picture
outlined above to make sense.
\par
This is obtained in Matrix string theories (DVV)~\cite{motl,dvv} by use of
supersymmetry.
In the limit of vanishing string coupling, the adjoint scalars $X_M$ and fermions
$\Theta_\alpha$ of the theory
commute with each other and it is possible to choose a gauge in which
they are all diagonal. However no Vandermonde determinant arises
from the Faddeev-Popov procedure because contributions from  $X_M$ and
$\Theta_{\alpha}$ cancel exactly.
\par
In~\cite{us,uscorfu} we showed that the same cancellation of Vandermonde
determinants occurs in pure two-dimensional Yang-Mills theory on a torus or a
cylinder if the gauge where the field strength $F$ is diagonal is chosen.
The ensuing theory has matrix string states, just as the DVV model, and its
partition function is closely related to the one of the DVV model 
itself~\cite{kv,Grignani:1999sp}.
In this paper we show that the results of~\cite{us,uscorfu} can be extended
to include Riemann surfaces of arbitrary genus and arbitrary number of
borders, if one considers a particular generalized Yang-Mills theory (in the
sense of~\cite{dls94,gsy95}) which has the property to coincide with
ordinary YM2 on flat surfaces. For the generalized theory it is shown that
the cancellation of the Vandermonde determinants occurs also on arbitrary
Riemann surfaces and that the theory is consistently described
by a string theory with a U$(1)$ gauge symmetry on its world-sheet\footnote{%
Also in the case of Matrix string theory there is a remaining  
U$(1)$ gauge symmetry on the world-sheet whose role was emphasized in
\cite{Bonelli:1998yt}.}.
\par
\section{Generalized Yang-Mills theory in the unitary gauge}
We begin by considering the partition function of a generalized YM2  on an
arbitrary Riemann surface~\cite{wit92,dls94,gsy95}. We shall then proceed to
fix the gauge in which $F$ is diagonal  by following essentially 
Ref.~\cite{blth}, where the partition function for ordinary YM2 was calculated
in the same gauge.
The partition function is given by
\begin{equation}
Z({\cal M})=\int {\cal D}A {\cal D}F\exp\left\{- \int_{\cal M}
d\mu\, V(F)+ \ii\, \tr \int_{\cal M} f(A) F \right\}~,
\label{partriemann}
\end{equation}
where $d \mu$ is the volume form on $\cal M$ and $f(A) $ is given by
\begin{equation}
f(A)= d A -\ii A\wedge A~.
\label{effea}
\end{equation}
In Eqs. (\ref{partriemann}) and (\ref{effea}) $F$ is a $N \times N$ hermitian
matrix and $A$ is a one-form on $\cal M$ with values in the space of
hermitian matrices. $V(F)$ denotes an arbitrary gauge invariant potential,
namely a potential that depends only on the eigenvalues of $F$, with in
principle an arbitrary dependence from the two-dimensional metric of the
surface.
In order to reproduce ordinary YM2 it is enough to choose
\begin{equation}
V(F) = \frac{g^2_{\rm YM}}{2} \tr F^2
\label{pot}
\end{equation}
and perform the quadratic functional integral in $F$.
\par
Our gauge choice consists in conjugating the $N\times N$ hermitian
matrix  $F$ into a diagonal form, namely into its Cartan sub-algebra.
As discussed in the previous section, this can always be done locally,
according to Eq. (\ref{Pgaugefix}).
The gauge fixed action, including the appropriate Faddeev-Popov ghost term,
can be written as the sum of two terms:
\begin{equation}
S_{\rm BRST}({\cal M},t) = S_{\rm Cartan} + S_{\rm off-diag}~,
\label{action}
\end{equation}
where $S_{\rm Cartan}$ involves the diagonal part of $A_{\mu}$ and exhibits
the residual U$(1)^N$ gauge invariance:
\begin{equation}
S_{\rm Cartan}= \int_{\cal M}
\left[ V(\lambda)d\mu -\sum_{i=1}^N
\ii \lambda_i d A^{(i)} \right]~,
\label{Scartan}
\end{equation}
where $A^{(i)}$ is the $i$-th diagonal term of the matrix form $A$.
The Faddeev-Popov ghost term and the off-diagonal part of $A$ are contained in
$S_{\rm off-diag}$ which can be cast into the following form:
\begin{equation}
S_{\rm off-diag} =  \int_{\cal M} d\mu \sum_{i>j}  (\lambda_i -
\lambda_j)\left[  \frac{1}{2} \epsilon^{ab} \hat{A}_a^{ij}\hat{ A}_b^{ji}
+ \ii ( c^{ij} \bar{c}^{ji}+ \bar{c}^{ij} c^{ji}) \right]~,
\label{offdiag}
\end{equation}
where $ \hat{A}_{a}^{ij} $
is given by
\begin{equation}
\hat{A}_{a}^{ij} = E_a^{\mu} A_{\mu}^{ij}
\label{flatA}
\end{equation}
and  $E_a^{\mu}$ denotes the inverse of the two-dimensional vierbein. The fields
$c^{ji}$ and $\bar{c}^{ij}$ are respectively the ghost and anti-ghost fields
associated to the gauge condition $F^{ij}=0$.
\par
The action  (\ref{offdiag}) contains the
same number of fermionic and bosonic degrees of freedom and it is invariant,
for each value of the composite index $[ij]$, with respect to a set of symmetry
transformation with Grassmann-odd parameters which are reported in~\cite{us}.
One would expect, as a result of these ``supersymmetries",
a complete cancellation
of the bosonic and fermionic contributions in the partition function.
However this supersymmetry is broken  on a generic
Riemann surface by the measure of the functional integral.
The reason is that the supersymmetric partners of the ghost and anti-ghost
fields are the zero-forms $\hat{A}_{a}^{ij}$, which are the components of the
one-form $A$ in the base of the vierbein (\ref{flatA}). So if the measure of the
functional integral contains  the one-form $ A^{ij}$, namely is
${\cal D}A_{\mu}^{ij} {\cal D}c^{ij}{\cal D}\bar{c}^{ij}$,
there is  a mismatch in the number
of bosonic and fermionic degrees of freedom as on a curved surface the
``number'' of zero-forms and  one-forms does not coincide.
This results into an anomaly that was explicitly calculated in~\cite{blth}:
\begin{equation}
\int \prod_{i>j} {\cal D}c^{ij}{\cal D}\bar{c}^{ij}{\cal D}A_{\mu}^{ij}
e^{-S_{\rm off-diag}}
= \exp \left\{{1 \over 8 \pi} \int_{\cal M} d\mu R \sum_{i>j} \log |\lambda_i -
\lambda_j| \right\}~.
\label{anomaly}
\end{equation}
\par
After integration over the non-Cartan fields, the theory becomes purely abelian
with a U$(1)^N$ gauge symmetry. Taking into account (\ref{anomaly}), it is
described by the effective action
\begin{equation}
S_{\rm eff}= \int_{\cal M}
\left[ \tilde{V}(\lambda)d\mu -\sum_{i=1}^N
\ii \lambda_i d A^{(i)} \right]~,
\label{Scartan2}
\end{equation}
where $\tilde{V}(\lambda)$ is given by
\begin{equation}
\tilde{V}(\lambda) = V(\lambda) - \frac{1}{8 \pi} R \sum_{i>j} 
\log |\lambda_i -\lambda_j|~.
\label{potential}
\end{equation}
\par
The anomaly term at the r.h.s. of (\ref{potential}) is divergent when two
eigenvalues coincide and requires some regularization procedure.
In  \cite{blth}, the calculation of the functional integral for the
U$(1)^N$ theory was done and it turned out to be equivalent to replacing each
eigenvalue $\lambda_i(x)$ with an arbitrary constant  integer value:
$\lambda_i(x) \rightarrow n_i $, where $n_i$ is the flux of the abelian field
strength. In this way the anomaly term
correctly reproduces the dependence of the standard partition 
function~\cite{r90,w91,bt92} from the genus $G$ of ${\cal M}$, namely the factor
$\prod_{i>j}(n_i - n_j)^{2 - 2G}$. However while in other gauges~\cite{bt92}
and in the lattice formulation the integers $n_i$, being labels of irreducible
representations of U$(N)$, satisfy the condition $n_1>n_2>n_3>...>n_N$, in the
unitary gauge nothing seems to forbid the existence of ``non regular
terms"\footnote{We follow the terminology of Ref.~\cite{blth}}, namely of terms
with at least two such integers coinciding.
\par
Non regular terms, which are divergent for $G>1$, were eliminated by a rather
{\it ad hoc} regularization procedure in~\cite{blth}, leaving the problem of a
completely consistent derivation of the standard partition function of YM2 in
the unitary gauge still open.
It should be noticed that all non trivial maps  $\Pi_1(x|{\cal M})
\rightarrow S_N$ lead
to sectors made entirely by non regular terms; in fact in a non trivial sector
at least two eigenvalues must  belong to the same connected world-sheet and,
being constant,  have to coincide. So any regularization scheme
that sets to zero the non regular terms also disregards all sectors
corresponding to non trivial maps  $\Pi_1(x|{\cal M})\rightarrow S_N$ .
\par
As already pointed out in~\cite{us}, non regular terms are finite in the 
case of the torus ($G=1$), and for any surface with zero curvature
the anomaly term at the r.h.s. of Eq. (\ref{potential}) vanishes. In this case
no regularization is needed and there is no reason to neglect the non trivial
maps in $\Pi_1(x|{\cal M}) \rightarrow S_N$, which for $G=1$ are labeled by
pairs $(Q_1,Q_2)$ of commuting permutations.
The calculation was done by the authors of~\cite{us}, initially with the hope
that the sum over the non trivial maps would just amount to cancel the non
regular terms. Surprisingly this happens only if the sum is restricted to the
subset of sectors with say $Q_2={\bf 1}$, while the entire sum leads to a
partition function~\cite{us} that has a wider spectrum than the standard one,
and which is closely related to the partition function of the DVV string matrix
model~\cite{kv,Grignani:1999sp}.
\paragraph{Higher genus surfaces and coupling to gravity}
In this paper we take a stronger point of view and  argue that
non-trivial sectors can be consistently kept also on space-times ${\cal M}$ of 
genus $G>1$. Our point of view  can be summarized as
follows:
\begin{enumerate}
\item
Quantization in the unitary gauge beyond the $G=1$ case would ask for a 
consistent regularization scheme (yet to be found) to cancel the logarithmic 
singularities in the anomaly term at the  r.h.s. of Eq. (\ref{potential}).
We propose instead to achieve the same result by modifying the
potential  $V(\lambda)$ to include in it a term that cancels exactly the 
divergent anomaly term so that the effective potential $ \tilde{V}(\lambda)$ 
is regular and of the form
\begin{equation}
\tilde{V}(\lambda)=\sum_{i=1}^N  v(\lambda_i)~,
\label{regpot}
\end{equation}
with
\begin{equation}
v(\lambda) =\sum_{k=1}^{\infty} t_k \lambda^k~.
\label{regpot2}
\end{equation}
In this way we define a generalized YM2 theory whose action reduces to
the one of standard YM2 in the limit of zero curvature and with a potential 
$v(\lambda) = \frac{g^2_{\rm YM}}{2} \lambda^2$.
In other words: we consider a generalized YM2 theory which is the
ordinary YM2 with a non standard coupling to gravity.
\item
It is possible, in our opinion, that the one presented above is the only
consistent way to quantize YM2  in the unitary gauge.
If that is the case the unitary gauge is
not a mere gauge choice, but it defines an entirely new theory.
Some possible reasons for it are discussed below.
\end{enumerate}
The redefinition of the potential exactly corresponds to the cancellation of
the anomaly. So it is natural to ask: is it possible to define the theory
from the beginning in such a way that the fermionic symmetry of (\ref{offdiag})
is not broken by the integration measure?
The obvious answer is to use the zero-forms
$\tilde{A}^{ij}_a$ rather than $A^{ij}_{\mu}$ as the independent degrees of
freedom in the functional integration measure, which becomes
${\cal D}\tilde{A}^{ij}_a{\cal D}c^{ij}{\cal D}\bar{c}^{ij}$.
This amounts to consider the gauge fields as defined in the tangent space, like
spinors, when space-time is curved. In spite of the fact that Eq. (\ref{flatA})
is locally invertible this results into a different counting of degrees of
freedom. This is apparent in a lattice formulation. Consider a 
discretized manifold made of
squares with sides all of the same length. The ghost and anti-ghost fields sit
on the sites of the lattice and the gauge fields on the links. On a site $j$
where the coordination number $c_j$ is 4 (zero curvature) the gauge degrees of
freedom match exactly the number of degrees of freedom of the ghost anti-ghost
system, otherwise the mismatch is proportional to $c_j-4$, namely to the
curvature. This argument provides a lattice based derivation of Eq.
(\ref{anomaly}). With the gauge fields defined on the tangent space, instead,
the degrees of freedom at each site coincide with the one in flat
space irrespective of the curvature and no anomaly arises.
\par
We do not know yet how to formulate on a lattice (for instance in the framework
of Regge calculus) the coupling to gravity of gauge fields associated to the
tangent space. On the other hand, there are indications that ordinary lattice
gauge theory
might be inadequate in curved space-time. Consider for instance the standard
Wilson action: the plaquette variable $U_{\rm pl}$ carries no information about
the metric and hence its continuum limit can only depend on quantities that can
be written as differential forms, namely
\begin{equation}
U_{\rm pl} \sim {\bf 1} + F_{\mu\nu} dx^{\mu} dx^{\nu}~.
\label{plaq}
\end{equation}
However the Lagrangian density $F_{\mu\nu} F_{\rho\sigma} g^{\mu\rho}
g^{\nu\sigma} \sqrt{g}$ has a non trivial dependence on the metric and it can
be written in the language of differential forms only by using the
Hodge dual operator.   Therefore the coupling of gauge theories to gravity
on a lattice seems to require, instead of the traditional Wilson action, a
first order formalism\footnote{The two-dimensional case is an exception from
this point of view: in this case the action depends on the metric only through
the area, and the Wilson action can be used provided the correct volume element
is used} (see for instance~\cite{cdm1}) where the plaquette variable is coupled
with an auxiliary field which in generic space-time dimensions is an
antisymmetric tensor $F_{ab}$, that is a two-form, in the tangent space indices.
\paragraph{Effective action}
With the effective potential $\tilde{V}(\lambda)$
given by Eqs. (\ref{regpot}) and
(\ref{regpot2}) the
action (\ref{Scartan2}) reduces to the sum of $N$ decoupled U$(1)$ theories,
namely
\begin{equation}
S_{eff} = \sum_{i=1}^{N} \int \left( v(\lambda_i) - \ii \lambda_i
dA^{(i)} \right)~.
\label{u1act}
\end{equation}
\par
In the functional integral, however, one has to sum over all sectors corresponding
to the homomorphisms $\Pi_1(x|{\cal M}) \rightarrow S_N$, as discussed in the
previous section.
So in a non trivial sector the U$(1)$ theories in (\ref{u1act}) are coupled by
the boundary conditions:
\begin{equation}
\gamma~~:~~\lambda_i(x) \rightarrow \lambda_{P_{\gamma}(i)}(x)~,
\label{bc}
\end{equation}
where $\gamma$ is an element of $\Pi_1(x|{\cal M})$ and $P_{\gamma}$ the
corresponding element of $S_N$ in the homomorphism. As a result, rather than
$N$ decoupled U$(1)$ theories, we have a U$(1) $ theory on each connected part
of the world-sheet, whose area is an integer multiple $k$ of the area ${\cal A}$
of the target space.
To each connected world-sheet is associated a U$(1)$ partition function,
given by:
\begin{equation}
Z_{{\rm U}(1)}(k {\cal A}) = \sum_{n=-\infty}^{+\infty}
{\rm e}^{-k {\cal A} v(n)}~,
\label{u1pf}
\end{equation}
where $v(n)$ is  given by Eq. (\ref{regpot2}).
Given a covering of ${\cal M }$ with $s_k$ ($\sum_k k s_k=N$) connected parts 
of area $k {\cal A}$, its contribution to the partition function is then given
by
\be
Z_{\{s_k\}} = \prod_{k=1}^{\infty} \left[Z_{{\rm U}(1)}(k {\cal A})
\right]^{s_k}~.
\label{essekappa}
\ee
The total partition function is then obtained by summing  (\ref{essekappa})
over all coverings.
\par
If the surface $ {\cal M}$ has boundaries, the states on the $A$-th boundary
$\gamma_A$ are characterized by the conjugacy class of
$P(\gamma_A)$, namely by the lengths of the cycles of  $P(\gamma_A)$, and by
the U$(1)$ holonomies for each cycle of  $P(\gamma_A)$. If $\theta$ is the sum
of the U$(1)$ holonomies of all cycles on a given connected world-sheet, the
U$(1)$ partition function (\ref{u1pf}) is replaced by the kernel:
\begin{equation}
K_{{\rm U}(1)}(k{\cal A},\theta) = \sum_{n=-\infty}^{+\infty}
{\rm e}^{-k {\cal A} v(n) +
\ii n \theta}.
\label{u1k}
\end{equation}
The kernel associated to a surface with an arbitrary
number of boundaries is
obtained by summing over all coverings with the prescribed conjugacy classes of
$S_N$ on the boundaries, each covering being weighted with the product of the
U$(1)$ kernel functions (\ref{u1k}) associated to its connected parts.
\par
In conclusion, after integration on the non-Cartan degrees of freedom, the
resulting effective theory is a string theory (theory of coverings without
branch points) with a U$(1)$ gauge theory on the world-sheet. In the next
section we shall show that this
can be described as a lattice gauge theory, with
a gauge group ${\cal G}_N$ which is the semi-direct product of $S_N$ and U$(1)^N$.
\section{ Lattice formulation of the effective theory}
Let us consider a surface $\cal M$ with the topology of a disc.
Since the homotopy group of the disc is trivial, only the identical
permutation can be associated to the boundary of the disc and the
only covering consists of $N$ disconnected copies of the disc.
According to the previous discussion the kernel on a disc is then
given by
\begin{equation}
K\left((P,\varphi), {\cal A}\right) = \delta(P)\sum_{n_i}
\exp\left\{\sum_{i=1}^{N}
\left(\ii n_i \varphi_i-{\cal A} v(n_i)\right)\right\}~,
\label{boltzmann}
\end{equation}
where ${\cal A}$ is the area of the disc and $\varphi_i$ the invariant angles of
the U$(1)$ holonomies. The kernel is vanishing for any permutation $P$ other
than the identity.
Consider now the group ${\cal G}_N$ defined as the subgroup of U$(N)$ given by
the matrices of the form
\begin{equation}
(P,\varphi)={\rm diag}\left(e^{\ii\varphi_1},\dots,e^{\ii\varphi_N}\right)\;P~,
\label{def}
\end{equation}
where $\varphi_k$ are real and $P$ is a permutation matrix:
\begin{equation}
P_{ij}=\delta_{iP(j)}~.
\end{equation}
Note that with this definition we have $(PQ)_{ij}=\delta_{i\;P\circ Q(j)}$,
while on a $N$-vector $\varphi$ we have $(P\varphi)_i=\varphi_{P^{-1}(i)}$.
The product in ${\cal G}_N$ is the ordinary matrix product: in the notation of
Eq. (\ref{def}) the group product reads
\begin{equation}
(P,\varphi)\;(Q,\theta)=(PQ, \varphi+P\theta)~,
\label{product}
\end{equation}
where it is clear that ${\cal G}_N$ can be viewed as a semi-direct product of 
$S_N$ and U$(1)^N$. The inverse of a group element  is given by
\begin{equation}
(P,\varphi)^{-1}=(P^{-1}, -P^{-1}\varphi)~.
\label{inverse}
\end{equation}
The kernel on a disc $K\left((P,\varphi), {\cal A}\right)$ is invariant under
gauge transformations of ${\cal G}_N$. In fact from
\begin{equation}
(Q,\theta)(P,\varphi)(Q,\theta)^{-1}=(QPQ^{-1},
\theta+Q\varphi-QPQ^{-1}\theta)
\label{conjugacy}
\end{equation}
it follows almost immediately that
\begin{equation}
K\left((Q,\theta)(P,\varphi)(Q,\theta)^{-1}, {\cal A}\right) =
K\left((P,\varphi), {\cal A}\right)~.
\label{inv}
\end{equation}
This property will enable us to formulate the theory described in the previous
section as a gauge theory of ${\cal G}_N$ on a lattice. In fact suppose to draw on the
target space ${\cal M}$ a lattice and to associate to each link $\alpha$ an
element $(P_{\alpha},\varphi_{\alpha})$ of ${\cal G}_N$. The product of all the
elements of ${\cal G}_N$ along a plaquette defines the plaquette variable
$(P_{\rm pl},\varphi_{\rm pl})$ and the partition function is defined as
\begin{equation}
Z_{\rm latt} = \prod_{\alpha}\sum_{P_{\alpha}} \int_0^{2 \pi} d\varphi_{\alpha}
\prod_{\rm pl} K\left((P_{\rm pl},\varphi_{\rm pl}), {\cal A}_{\rm pl}
\right)~.
\label{lattpart}
\end{equation}
As in the case of the heat-kernel action for a U$(N)$ gauge theory, the
Boltzmann weight for a single plaquette is left invariant in form if one sews
two plaquettes together by integrating over their common link.  In fact a
straightforward calculation gives:
\ba
&&\sum_{Q\in S_N}\int \prod_i \frac{d\theta_i}{2\pi} K\left((P_1,\varphi_1)
(Q,\theta),{\cal A}_1\right) K\left((Q,\theta)^{-1}(P_2,\varphi_2), {\cal A}_2
\right)\nonumber\\
&&\;\; =K\left((P_1, \varphi_1)(P_2, \varphi_2), {\cal A}_1+{\cal A}_2\right)~.
\label{glue}
\ea
This property insures that the result obtained on a lattice is valid
also in the continuum limit. On the other hand, it also allows to solve exactly
the theory by reducing the surface to a single plaquette with suitably
identified links. Again this is in complete analogy with the solution of YM2
obtained in~\cite{r90}.
Let us consider first the case of the torus, namely of a square with identified
opposite sides. The partition function is given by:
\ba
Z_N({\cal A})&=&
\sum_{P,Q\in S_N}\int\prod_i\frac{d\varphi_i d\theta_i}{4\pi^2}
K\left((P,\varphi)(Q,\theta)(P,\varphi)^{-1}(Q,\theta)^{-1},{\cal A}\right)
\nonumber\\
&=&\sum_{P,Q\in S_N}\int\prod_i\frac{d\varphi_i d\theta_i}{4\pi^2}
\nonumber\\
&& K\left( \left(PQP^{-1}Q^{-1}, \varphi+P\theta-PQP^{-1}\varphi
-PQP^{-1}Q^{-1}\theta\right),{\cal A}\right)\nonumber\\
&=&\sum_{P,Q\in S_N}\delta(PQP^{-1}Q^{-1})
\int\prod_i\frac{d\varphi_i d\theta_i}{4\pi^2}
\nonumber\\
&& \sum_{\{n_i\}}
\exp\left\{\sum_i\left(\ii n_i
\left[\phi_i+(P\theta)_i-(Q\phi)_i-\theta_i)\right]
-{\cal A} v(n_i)\right)\right\}~.
\ea
Therefore the partition function is expressed as a sum over pairs of
commuting permutations, that is over unbranched $N$-coverings of the torus.
The integration over $\theta$ and $\phi$ produces respectively a
 factor $\delta_{n_i\;n_{P^{-1}i}} $ and $\delta_{n_i\;n_{Q^{-1}i}} $.
The result of the integration is to equate all the $n_i$'s
belonging to a given connected component of the covering defined by
$(P,Q)$. A U$(1)$ partition function given in (\ref{u1pf}) is associated to
every such connected component of area $k {\cal A}$, which is a torus with area
$k {\cal A}$.
This result is in complete agreement both with the prescription given at the end
of the previous section and with the results of~\cite{us}.
\begin{figure}
\centering
\includegraphics[width=0.56\textwidth]{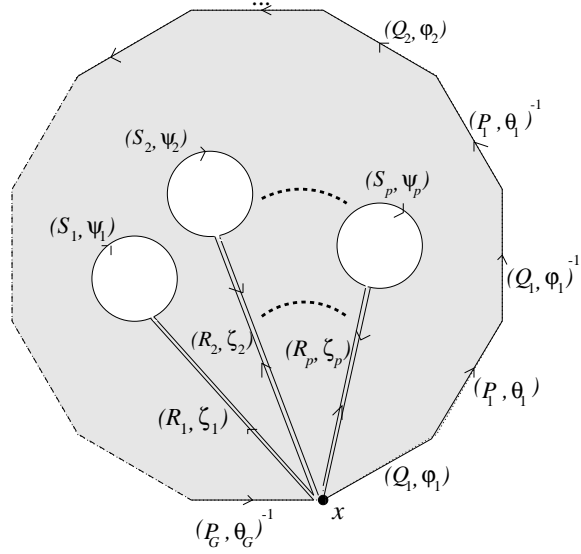}
\caption{The lattice theory on a surface of genus $G$ with $p$ boundaries
can be reduced to a single plaquette with suitably identified edges.}
\label{fig:genusG}
\end{figure}
\par
These results can easily be generalized to a surface of arbitrary genus and an
arbitrary number of boundaries.
In Fig. \ref{fig:genusG} a surface of genus $G$ with $p$ boundaries
is represented as a single plaquette with suitably identified edges. The group
elements of ${\cal G}_N$ associated to each link are also indicated in the figure.
The partition function is then given by:
\ba
&&Z_N\left(\widehat{S_1\psi_1},\widehat{S_2\psi_2},....,\widehat{S_p\psi_p},
{\cal A},G\right)= \sum_{P_a,Q_a,R_{A}}\int \prod_{i=1}^N\,
\prod_{a=1}^{G} \frac{d\varphi_{a,i} d\theta_{a,i}}{4 \pi^2}\prod_{A=1}^p
 \frac{d\zeta_{A,i}}{2 \pi}\nonumber\\
&&\delta(\Omega) \sum_{\{n_i\}} \left\{\sum_i \{\ii n_i \omega_i - {\cal A}
v(n)\}\right\}~,
\label{partgen}
\ea
where the hats at the l.h.s. are to denote that $Z_G$ only depends on the
conjugacy classes of $(S_{A},\psi_{A})$.
$\Omega$ is the product of all permutations along the plaquette.
Using the shorthand notation
\begin{equation}
V_{A}=R_{A} S_{A}R_{A}^{-1}~,~~~~~~~~~~~W_a=Q_aP_aQ_a^{-1}
P_a^{-1}
\label{shorthand}
\end{equation}
it can be written as
\begin{equation}
\Omega= V_1 V_2 ...V_p W_1 W_2.......W_G~.
\label{omega}
\end{equation}
The angles $\omega_i$ are the corresponding U$(1)^N$ invariant angles:
\ba
\omega &=& \sum_{A=1}^p ( V_1 V_2
...V_{A-1}R_{A}\psi_{A} +
V_1V_2...V_{A-1}\zeta_{A} -
V_1V_2...V_{A}\zeta_{A} )\nonumber\\
&+&\sum_{a=1}^G
(V_1...V_pW_1...W_{a-1}\varphi_a - V_1...V_pW_1...W_{a-1}Q_aP_aQ_a^{-1}
\varphi_a \nonumber\\
&+& V_1...V_pW_1...W_{a-1}Q_a\theta_a
-V_1...V_pW_1...W_{a-1}W_a\theta_a)~.
\label{angletot}
\ea
The integrations over $\zeta_{A,i}$, $\varphi_{a,i}$ and $\theta_{a,i}$
produce a set of delta functions that identify some (or all) of the integers
$n_i$, namely:
\ba
\widehat{\delta(n)} &=&\prod_{i=1}^{N}\prod_{A=1}^p
\delta \left( n_i,n_{V_1V_2..V_{A-1}V_{A}V_{A-1}^{-1}...
V_2^{-1}V_1^{-1}(i)}\right)\nonumber \\
&\times& \prod_{i=1}^{N}
\prod_{a=1}^G \delta\left(n_i,n_{V_1...V_pW_1...W_{a-1}Q_aP_aQ_a^{-1}
W_{a-1}^{-1}...W_1^{-1}V_p^{-1}...V_1^{-1}(i)}\right)\nonumber\\
&\times& \prod_{i=1}^{N}\prod_{a=1}^G \delta \left(n_i,n_{V_1...V_pW_1...W_{a-1}
Q_aP_aQ_a^{-1}P_a^{-1}Q_a^{-1}W_{a-1}^{-1}...W_1^{-1}V_p^{-1}...V_1^{-1}(i)}
\right)~.\nonumber\\
\label{deltas}
\ea
The set of $\delta$-functions of Eq. (\ref{deltas}) are in one-to-one
correspondence with the generators of the homotopy group of ${\cal M}$
with base point $X$. So {\it all the integers $n_i$ with indices that can
obtained from each other by acting with an element of the homotopy group are
identified}.
\par
By taking into account the definitions (\ref{shorthand}) and (\ref{omega}) and
the result (\ref{deltas}), the final expression for the partition function
(\ref{partgen}) is given by
\ba
&&Z_N\left(\widehat{S_1\psi_1},\widehat{S_2\psi_2},....,\widehat{S_p\psi_p},
{\cal A},G \right)= \sum_{P_a,Q_a,R_{A}}
\delta(\Omega) \sum_{\{n_i\}} \widehat{\delta(n)}\nonumber \\
&&\times
\exp \left\{\sum_i \left( \ii n_i \sum_{A=1}^p ( V_1 V_2...V_{A-1}
R_{A}\psi_{A})_i - {\cal A} v(n_i)\right)\right\}~.
\label{partgen2}
\ea
In Eq. (\ref{partgen2}) the sum over $P_a,Q_a,R_{A}$ with the constraint given
by $\delta(\Omega)$ counts the number of coverings with the prescribed
conditions at the boundaries, while $\widehat{\delta(n)}$ ensures that each
connected world-sheet contributes with a factor which is exactly the U$(1)$
kernel given in (\ref{u1k}).
It should be noticed that the r.h.s. of (\ref{partgen2}) depends only on the
conjugacy classes of $(S_{A}, \psi_{A})$, namely on the cycle
decomposition of $S_{A}$ and, for each cycle $C$ of $S_{A}$, on the
quantities $\psi_{A}^{C}= \sum_{i\in C} \psi_{A,i}$.
This is a consequence of gauge invariance under ${\cal G}_N$ transformations
but it can be seen explicitly from (\ref{partgen2}) by noticing that
$S_{A}$ appears only in the expression $R_{A} S_{A}
R_{A}^{-1}$ (with a sum over $R_{A}$), and that all $\psi_{A,i}$
with $i\in C$ are coupled to the same integer $n$ due to the effect of
$\widehat{\delta(n)}$.
Finally it should be remarked that the kernel (\ref{partgen2}) is completely
consistent in the sense that it is reproduced by sewing two surfaces together
according to the standard procedure, namely by identifying the conjugacy classes
of the boundaries to be sewn and by integrating over the group.
\section{Partition function on arbitrary genus}
In order to write an explicit expression for the partition function (or kernel)
given in (\ref{partgen2}), we have to use techniques developed by several
authors~\cite{gt,ksw}  that allow to write the number of coverings on a given
surface in terms of the characters of $S_N$. In this section we will restrict
the discussion to the case of surfaces without boundaries, while in the next
section the general case of surfaces with boundaries will be discussed.
We can proceed in four steps:
\begin{enumerate}
\item
Write the partition function that  describes the statistics of the coverings
without branch points on a surface of genus $G$, that is
\begin{equation}
Z^{(\rm cov)}(G,q) = \sum_{k=1}^{\infty} Z_k^{(\rm cov)}(G) q^k~,
\label{Zcov}
\end{equation}
where $Z_k^{(\rm cov)}(G)$ is the number of $k$-coverings. This partition
function was studied in~\cite{gt,ksw} and it will be given below.
\item
Write the partition function $F^{(\rm cov)}(G,q)$ that describes the statistics
of {\it connected} coverings. This is obtained by taking the logarithm of
$Z^{(\rm cov)}(G,q)$:
\begin{equation}
F^{(\rm cov)}(G,q) = \log Z^{(\rm cov)}(G,q) = \sum_{k=1}^{\infty}
F_k^{(\rm cov)}(G) q^k~.
\label{Fcov}
\end{equation}
\item
Associate to each connected covering the U$(1)$ partition function (\ref{u1pf}):
\begin{equation}
F({\cal A},G,q) = \sum_{k=1}^{\infty} F_k^{(\rm cov)}(G) \sum_{n=-\infty}^{+\infty}
{\rm e}^{-k {\cal A}v(n)} q^k = \sum_{n=-\infty}^{+\infty}
F^{(\rm cov)}(G,{\rm e}^{-{\cal A} v(n)} q)
\label{freen}
\end{equation}
\item
The exponential\footnote{In~\cite{us} for the case $G=1$
both signs in ${\rm e}^{\pm F(G,q)}$ were considered, the minus sign corresponding
to a fermionic model. In the general case $G>1$ this choice seems to
be inconsistent. In fact this corresponds to weighting with a factor $(-1)$ each
connected world-sheet, and we could not find any prescription for sewing surfaces
together that preserve this property.} of $F({\cal A},G,q)$ is then the
generating function (i.e. grand-canonical partition function) for the partition
functions given in (\ref{partgen2}) with no boundaries ($p=0$):
\begin{equation}
Z({\cal A},G,q) = {\rm e}^{ F({\cal A},G,q)} =
\sum_{N=1}^{\infty} Z_{N}({\cal A},G)
q^N =\prod_{n=-\infty}^{+\infty} Z^{(\rm cov)}(G,{\rm e}^{-{\cal A} v(n)} q)~.
\label{pf}
\end{equation}
\end{enumerate}
\par
All we need to know, in order to write $Z({\cal A}G,q)$, is the explicit 
expression of $Z^{(\rm cov)}(G,q)$, that was given in \cite{gt,ksw} as a sum
over irreducible representations of $S_N$ as $N \rightarrow \infty$:
\begin{equation}
Z^{(\rm cov)}(G,q) =\sum_{\{h_i\}} \left[ N^{|h|} \Lambda(h) \right]^{2G-2}
q^{|h|}~,
\label{Zcov2}
\end{equation}
where
\begin{equation}
|h| = \sum_{i=1}^N h_i - \frac{1}{2}N(N-1)
\label{hmod}
\end{equation}
and
\begin{equation}
\Lambda(h)=\frac{\prod_{i=1}^N h_i!}{\prod_{i<j}(h_i-h_j)}~.
\label{Lambda}
\end{equation}
The integers $h_i$ are subjected to the constraint $h_1>h_2>.....>h_N\geq0$ and
are related to the lengths $m_1,m_2,...,m_N$ of the rows of the corresponding
Young tableau by $h_i=N-i+m_i$. The quantity $|h|=\sum_i m_i$ is the number of
boxes in the Young tableau.
\par
The integer $N$ in Eq. (\ref{Zcov2}) is just the order of the matrices in the
complex matrix model introduced in Ref.~\cite{ksw}, and whose solution is given
by Eq. (\ref{Zcov2}). No such quantity is present in the original problem of
counting coverings, so there must be no dependence on $N$ at the end. The factor
$N^{|h|(2G-2)} = N^{2g-2}$ just counts the genus $g$ of the world-sheet and it
will be dropped in what follows. All the other quantities in (\ref{Zcov2})
depend only on the Young tableau and not on the order $N$ of the symmetric
group. However the limit $N \rightarrow \infty$ has to be taken in order to take
into account Young tableaux with an arbitrarily large number of rows.
\begin{figure}
\centering
\includegraphics[width=0.9\textwidth]{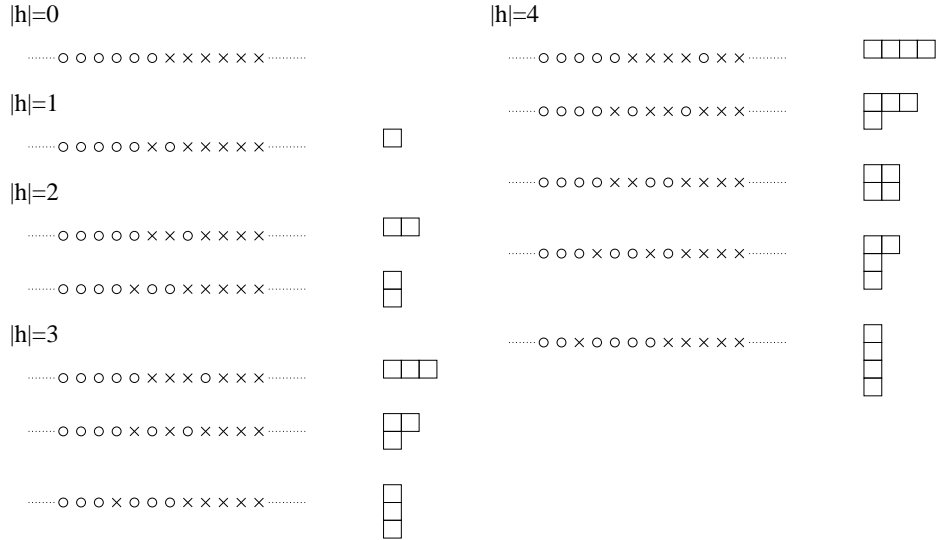}
\caption{Ground state and lowest-lying excited states in the fermionic
description of the representations of $S_N$, together with the corresponding
Young tableaux.}
\label{fig:reps}
\end{figure}
\par
In order to make this more explicit and also to simplify the expression of
(\ref{Zcov2}) we introduce a notation where each Young tableau is associated to
an excited state of a gas of fermions whose energy levels are the relative
integers (see~\cite{doug93} for an analogous notation in the case of U$(N)$).
The ground state ($|h|=0$) and the
lowest excited states are shown in Fig. \ref{fig:reps}; they are
in one-to-one correspondence with the Young tableaux.
The correspondence is as follows: let the
infinite sequence of relative integers $h_1>h_2>h_3>.........$ represent the
occupied levels and the complementary  sequence $k_1<k_2<k_3<.......$ the empty
levels. By definition $\{h_i\}\cup\{k_j\} = {\bf Z}$ and
$\{h_i\}\cap\{k_j\}=\emptyset$.
The position of the origin is irrelevant, so the sequences $\{h_i\}$ and
$\{k_j\}$ are defined modulo an integer shift. The length $m_i$ of the $i$-th
row of the Young tableau is given by the number of empty levels {\it below}
$h_i$:
\begin{equation}
m_i = \# k_j:\, k_j < h_i
\label{mi}
\end{equation}
and similarly the length $r_j$ of the $j$-th column by the number of the
occupied levels {\it above} $k_j$:
\begin{equation}
r_j=\# h_i:\, k_j < h_i~.
\label{rj}
\end{equation}
The total number of boxes is
the number of pairs $\{k_j,h_i\}$ for which $h_i>k_j$:
\begin{equation}
|h|= \# \{k_j,h_i\}:\, k_j < h_i~.
\end{equation}
This coincides with the number of ``moves" required to reach the excited
state under consideration from the ground state, a move being defined as a jump
of a fermion one level up to occupy an empty level.
The Young tableaux with given $|h|$ label the irreducible representations
of $S_{|h|}$, and the factor $\Lambda(h)$ given in Eq. (\ref{Lambda}) is
related to the dimension $d_{\{h\}}$ of the corresponding representation by:
\be
\Lambda(h)=\frac{|h|!}{d_{\{h\}}}~.
\label{dimrap}
\ee
A simple calculation shows that $\Lambda(h)$  can be rewritten as
$\prod_{\{h_i>k_j\}} (h_i-k_j)$.The partition function of the
coverings can then be written without any reference to $N$ as:
\begin{equation}
Z^{(\rm cov)}(G,q) = \sum_{h_1>h_2>....}
\left[ \prod_{\{h_i>k_j\}}(h_i-k_j)\right]^{2G-2} q^{|h|}=\sum_{\{h\}}
\left[\frac{d_{\{h\}}}{|h|!}\right]^{2-2G} q^{|h|}~.
\label{Zcov3}
\end{equation}
The  coefficients of the expansion of $ Z^{(\rm cov)}(G,q)$
can be easily calculated
from (\ref{Zcov3}), and the first few of them are reproduced in the Appendix.
The grand-canonical partition function $Z({\cal A},G,q)$ can be obtained in
principle by inserting (\ref{Zcov3}) into (\ref{pf}) and the first few
coefficient of the expansion are also given in the Appendix.
Except for $G=0$ and $G=1$ no compact expression for $Z^{(\rm cov)}(G,q)$ or
$Z({\cal A},G,q)$ is available.
\par
For $G=0$ we have $Z^{(\rm cov)}(G=0,q)= {\rm e}^q$
and the U$(N)$ partition function is simply given by
\begin{equation}
Z_N({\cal A},G=0) =
\left( \sum_{n=-\infty}^{\infty} {\rm e}^{-{\cal A}v(n)} \right)^N~.
\label{Zsphere}
\end{equation}
This means that for the topology of the sphere our theory reduces to $N$ copies
of the corresponding abelian theory.
\par
As already discussed in~\cite{uscorfu}, $Z^{(\rm cov)}(G,q)$ can be written 
for $G=1$ as an infinite product, and $Z({\cal A},G=1,q)$ is given by:
\begin{equation}
Z({\cal A},G=1,q) = \prod_{n=-\infty}^{\infty} \prod_{k=1}^{\infty}\frac{1}
{ 1 - q^k {\rm e}^{-k{\cal A} v(n)} }~.
\label{toruspf}
\end{equation}
\section{Surfaces with boundaries}
Consider a surface ${\cal M}_{G,p}$ of genus $G$ and with $p$
boundaries. As discussed in the previous section, the original U$(N)$ gauge
theory is described by the $N$-coverings of ${\cal M}_{G,p}$ with a residual
U$(1)$ gauge theory on each world-sheet. The states at the $A$-th boundary are
then characterized by :
\begin{enumerate}
\item The conjugacy class of $P(\gamma_A)$, where $\gamma_A$ is any closed path
that can be deformed  into the boundary (see Section 2). Each conjugacy class
is given in terms of the length of the cycles of $P(\gamma_A)$,
namely of the set of integers $\{r_l^{(A)}\}$ with 
$\sum_{l=1}^N  l r_l^{(A)} = N$.
\item The U$(1)$ holonomy associated to each boundary
of any world-sheet, that is
to each cycle of $P(\gamma_A)$. The holonomy is given by an invariant angle
$\theta_{l,\alpha}^{(A)}$, where the label $\alpha=1,2,...,r_l^{(A)}$ enumerates
the cycles of the same length. Since the cycles of same length are
indistinguishable, we expect the wave function to be symmetric in
the corresponding angles.
\end{enumerate}
\par
Let us first consider the case where no U$(1)$ gauge theory is present, namely
the pure theory of coverings. The grand-canonical partition function that
describes the statistics  of the coverings on a surface with $p$ boundaries
and genus $G$ is defined by
\begin{equation}
Z^{(\rm cov)}(G,p, t_l^{(A)}) =
\sum_{\{r_l^{(A)}\}} Z^{(\rm cov)}(G,p, r_l^{(A)})
\prod_{A=1}^{p}\prod_{l=1}^{\infty} \left[t_l^{(A)}\right]^{r_l^{(A)}}~.
\label{Zconbordi}
\end{equation}
Here $Z^{(\rm cov)}(G,p, r_l^{(A)})$ is the number of coverings
with $r_l^{(A)}$ cycles of length $l$ on the $A$-th boundary.
Its explicit form
was given in~\cite{ksw} and with the notations of Section 5 it reads
\begin{equation}
Z^{(\rm cov)}(G,p, r_l^{(A)}) = \sum_{\{h_i\}}
\left[ \Lambda(h) \right]^{2G-2 +p}
\prod_{A=1}^p  \frac{\chi_{h}(r^{(A)}_l)}{\prod_l l^{r^{(A)}_l} r^{(A)}_l!}~,
\label{Zconbordi2}
\end{equation}
where $\chi_{h}(r^{(A)}_l)$ are the characters of $S_{|h|}$ if
$r^{(A)}_1+ 2 r^{(A)}_2+3 r^{(A)}_3+.....=|h|$ and zero otherwise.
It is important to notice that sewing two surfaces together
reproduces the same grand-canonical partition function through the orthogonality
properties of the characters, in exactly the same way as it happens in
the Rusakov partition function of YM2.
\par
The grand-canonical partition function for {\it connected} coverings is obtained
as usual by
\begin{eqnarray}
F^{(\rm cov)}(G,p, t_l^{(A)}) & = & 
\log Z^{(\rm cov)}(G,p, t_l^{(A)})\nonumber\\ 
& = &
\sum_{r_l^{(A)}} F^{(\rm cov)}(G,p, r_l^{(A)})
\prod_{A=1}^{ p}\prod_{l=1}^{\infty}\left[t_l^{(A)}\right]^{r_l^{(A)}}~.
\label{Fconbordi}
\end{eqnarray}
The number of connected coverings $F^{(\rm cov)}(G,p, r_l^{(A)})$ can in
principle be calculated from (\ref{Fconbordi}) and (\ref{Zconbordi}),
although no closed expression is available.
\par
Let us now introduce the U$(1)$ gauge theory. We are interested in the kernels
$Z(G,p,r_l^{(A)},\theta^{(A)}_{l,\alpha})$, namely in the partition function
with prescribed boundary conditions.
It is convenient to introduce the kernels in momentum space of
the angular variables $\theta^{(A)}_{l,\alpha}$, namely
\begin{eqnarray}
Z\left({\cal A},G,p,r_l^{(A)},n^{(A)}_{l,\alpha}\right) 
& = & \int_{0}^{2 \pi}
\prod_{A,l,\alpha} d\theta^{(A)}_{l,\alpha}
\exp \left\{ \sum_{A,l,\alpha}
\ii n^{(A)}_{l,\alpha} \theta^{(A)}_{l,\alpha} \right\}\nonumber\\ 
&&
Z({\cal A},G,p,r_l^{(A)},\theta^{(A)}_{l,\alpha})~.
\label{momspace}
\end{eqnarray}
In momentum space the boundary conditions are now completely specified by
giving $r_l^{(A)}$ and $n^{(A)}_{l,\alpha}$. However, since two cycles of the
same length and momentum are indistinguishable, we can label the state at
each boundary $i$ by the number $r^{(A)}_{n,l}$ of cycles of length $l$ and
momentum $n$:
\be
r^{(A)}_{n,l}=\sum_{l,\alpha} \delta_{n,n^{(A)}_{l,\alpha}}
\label{erreenne}~.
\ee
 So we shall write $Z\left({\cal A},G,p,r^{(A)}_{n,l}\right)$ instead of
$Z\left({\cal A},G,p,r_l^{(A)},n^{(A)}_{l,\alpha}\right)$.
The corresponding grand-canonical function can be written as
\begin{equation}
Z\left({\cal A},G,p,t^{(A)}_{n,l}\right) = \sum_{\{r^{(A)}_{n,l}\}}
Z\left({\cal A},G,p,r^{(A)}_{n,l}\right) \prod_{A=1}^p \prod_{l=1}^{\infty}
\prod_{n=-\infty}^{+\infty}  \left[t_{n,l}^{(A)}\right]^{r^{(A)}_{n,l}}~.
\label{Zmom}
\end{equation}
As usual we also introduce the grand-canonical function that corresponds to
the connected coverings:
\begin{equation}
F\left({\cal A},G,p,t^{(A)}_{n,l}\right) =
\log Z\left({\cal A},G,p,t^{(A)}_{n,l}\right)~.
\label{fmom}
\end{equation}
The structure of $F\left({\cal A},G,p,t^{(A)}_{n,l}\right)$ is determined by the
following considerations. A U$(1)$ partition function, as in Eq. (\ref{u1pf}):
\begin{equation}
K_{{\rm U}(1)}(k{\cal A},\theta) = \sum_{n=-\infty}^{+\infty}
{\rm e}^{- k{\cal A }v(n) +
\ii n \theta}
\label{u1part}
\end{equation}
is associated to each connected part of the world-sheet.
Here $k$ is the degree of the covering and
$\theta $ is the sum over all the invariant U$(1)$ angles associated to the
cycles of the world-sheet's boundaries. It is clear then that, due to the
angular integrations in Eq. (\ref{momspace}), all cycles
belonging to the same connected covering have the same momentum $n$
irrespective of the boundary on which they are situated. Moreover,
the total length $|r|_n$ of the cycles with a given momentum
$n$ is the same for all boundaries of ${\cal M}_{G,p}$:
\begin{equation}
\sum_{l} l r^{(A)}_{n,l} = |r|_n~.
\label{Nn}
\end{equation}
In conclusion, each term of $F\left({\cal A},G,p,t^{(A)}_{n,l}\right)$
is characterized by a unique momentum $n$, is weighted by the factor
${\rm e}^{- |r_n|{\cal A} v(n)}$
coming from the U$(1)$ kernel and is proportional to the number
$F^{(\rm cov)}\left(G,p,r^{(A)}_{n,l}\right)$ of connected coverings with
cycles of given lengths:
\begin{equation}
F\left({\cal A},G,p,t^{(A)}_{n,l}\right)= \sum_{n=-\infty}^{+\infty}
F^{(\rm cov)}\left(G,p,r^{(A)}_{n,l}\right) \prod_{A=1}^p \prod_{l=1}^{\infty}
\left[t_{n,l}^{(A)}\right]^{r^{(A)}_{n,l}} {\rm e}^{-|r|_n{\cal A} v(n)}~.
\label{fmom2}
\end{equation}
The expression for the kernel $Z\left({\cal A},G,p,t^{(A)}_{n,l}\right)$ can be
obtained by exponentiating both sides of Eq. (\ref{fmom2}) and it can be
written as an infinite product:
\begin{equation}
Z\left({\cal A},G,p,t^{(A)}_{n,l}\right) = \prod_{n=-\infty}^{+\infty} 
Z^{(\rm cov)}\left(
G,p,{\rm e}^{-l{\cal A} v(n)} t^{(A)}_{n,l}\right)
\label{Zmom2}
\end{equation}
Each term in the infinite product in Eq. (\ref{Zmom2}) is the grand-canonical
partition function for the coverings defined in Eq. (\ref{Zconbordi})
with $t^{(A)}_l$ replaced by ${\rm e}^{-l v(n)} t^{(A)}_{n,l}$.
This is the analogue of Eq. (\ref{pf}) for kernels with boundaries.
\par
Starting from eq. (\ref{Zmom2}) one can obtain the result, already known fron
the lattice formulation os Section 4, that by sewing together two surfaces we
obtain a kernel described again by Eq. (\ref{Zmom2}) for the resulting surface.
This follows from the fact that same property holds for the partition function 
of the coverings $Z^{(\rm cov)}\left(G,p,t^{(A)}_{l}\right)$. In fact when
sewing two boundaries
together the U$(1)$ invariant angles on the two sides have to be identified in
pairs (for cycles of the same length!) and integrated over. This gives a delta
function of the corresponding momenta, so that for each subspace of
given momentum $n$ the sewing and cutting is unaffected by the other subspaces
and its statistics coincides with the one of the pure covering theory.
\section{Conclusions and further developments}
The quantization of YM2 in the unitary gauge on an arbitrary Riemann
surface has led us to consider a different model, a generalized YM2 theory
whose effective theory, after integration over the non-Cartan component of
the gauge fields, is described by a string with a U$(1)$ gauge group on the
world-sheet. This theory coincides with a gauge theory on a lattice with gauge
group ${\cal G}_N$, the semi-direct product of the symmetric group
$S_N$ and U$(1)^N$.
It may be surprising that the partition function on an arbitrary genus surface
of our original U$(N)$ gauge theory can be calculated exactly within the gauge
theory of a group ${\cal G}_N$ which is a subgroup of the original gauge group.
However there are correlators of the original U$(N)$ gauge theory that cannot be
calculated in the framework of the effective theory. Consider for instance
correlators of Wilson loops, which depend on the non-Cartan components of the
U$(N)$ gauge fields. If the Wilson loops are non intersecting, such
components are irrelevant as the quadratic term in the off-diagonal action
(\ref{offdiag}) depends at each space-time point  on both space-time components
of $\hat{A}_a^{ij}$. For intersecting Wilson loops instead
this term produces an
interaction between different sheets of the covering
whose effect cannot be described in terms of
the pure ${\cal G}_N$ theory. The situation is somewhat similar to the one in
multi-matrix models, where the angular degrees of freedom can be integrated out
in the calculation of the partition function and of certain correlators but are
involved in a non trivial way in the most general correlators.
\par
Besides its relation with YM2, the ${\cal G}_N$ gauge theory is interesting
by itself, and it can generalized in two different directions that we shall
briefly mention here, while leaving the details for a future
publication~\cite{usfut}.
\par
The first type of extension is the one already considered in~\cite{ksw}
for the theory of coverings. It consists in allowing coverings with
branch points, namely in allowing the strings to split and join.
In the framework of Matrix string theory this was discussed in
~\cite{ghv,Bonelli:1998yt}.
A branch point can be
described as a boundary where all the U$(1)$ invariant angles are set to zero.
In this case, if the permutation associated to the boundary is trivial this
just becomes an ordinary point, otherwise it becomes a point where a more or
less complicated interaction amongst the sheets of the covering occurs.
The simplest case  consists in a single string splitting into two
strings or vice-versa, and it is
described by a permutation with only one non trivial cycle of order $2$.
>From this point of view  the results of Section 6 are all we need to introduce
branch points in our theory. However we want to consider, as in~\cite{ksw},
the limit  where a continuum distribution of branch points is introduced.
For the theory of pure coverings, that is for the gauge theory of $S_N$, 
this corresponds to replacing the plaquette action, which for the 
unbranched theory of coverings is just a delta function, with a sum over 
irreducible representations of $S_N$:
\be
\delta(P)=\sum_R  d_R \chi_R(P) \rightarrow \sum_R  d_R \chi_R(P)
{\rm e}^{-C_R}~.
\label{replacement}
\ee
Here $P$ is the permutation associated to the plaquette and $C_R$ is any
quantity that depends only on the representation $R$ and that
contains all the information about the couplings of the different types of
branch points (see~\cite{ksw} for details).
\par
In our ${\cal G}_N$ gauge theory the introduction of a continuum of branch 
points will amount to  replacing the plaquette action (\ref{boltzmann}) with a 
generic expansion in the characters of the irreducible representations of 
${\cal G}_N$.
The weight of each representation should contain the area ${\cal A}$ of the
plaquette at the exponent, to insure good gluing properties as in (\ref{glue}),
and will depend both on the potential $v(n)$ that appears in the original action
and on the couplings of the different types of branch points encoded in
$C_R$. These couplings are new free parameters of the theory which are not
present in the original Lagrangian. As they describe interactions amongst
different sheets, they would presumably correspond to singular terms in
the U$(1)^N$ gauge theory obtained after fixing the unitary gauge, possibly of
the same type as the ones that have been eliminated by the redefinition
of the potential.  
A different type of extension is obtained by considering a more general gauge
group ${\cal G}_{L,M}$, defined as the semi-direct product of
$S_L$ and U$(M)^L$.
The product law in ${\cal G}_{L,M}$ is explicitly given by
\be
(P,g_i)(Q,h_i)=(PQ,g_i h_{P^{-1}(i)})~,
\label{mult}
\ee
where $P$ and $Q$ are elements of $S_L$, $g_i$ and $h_i$ ($i=1,2,...,L$) are
elements of U$(M)$ and the products at the r.h.s. of (\ref{mult}) are
accordingly group products in $S_L$ and U$(M)$.
The gauge theory based on this group describes a theory of coverings
(with possibility of introducing branch points as above) with a U$(M)$ on
each world-sheet. This theory would include both the standard YM2 and the
${\cal G}_N$
theory described in this paper as particular cases. Its plaquette action
(without branch points) can be easily constructed by following the same line of
reasoning used to obtain Eq. (\ref{boltzmann}), and it reads
\be
K\left( (P,g_i),{\cal A}\right) = \delta(P) \prod_{i=1}^L \sum_{R^{(i)}}
d_{R^{(i)}} \chi_{R^{(i)}}(g_i) e^{-{\cal A} C_{R^{(i)}}}~,
\label{plach}
\ee
where $R^{(i)}$ are the representations of the $i$-th group U$(M)$ and are
labeled by a set of $M$ integers $\{n^{(i)}\}$.
The whole theory can now be constructed by sewing plaquettes together and it is
essentially obtained by replacing in the formulas of Section 4 the integer
momentum $n$ with the representations of U$(M)$ and the plane wave function
$e^{\ii n \theta}$ with the characters $\chi_{R^{(i)}}(g) $.
By allowing branch points in the string theories based on ${\cal G}_{L,M}$ 
we can construct a large class of string theories, that range from 
the pure U$(M)$
theory (standard YM2 theory) at one end to the pure theory of coverings at the
other end. The well known result of Gross and Taylor~\cite{gt}, namely that in
the large $N$ YM2 is equivalent to a theory of coverings with a
set of allowed interactions, suggests that the same kind of dual description
may be possible also for the more general theories we have discussed.
\paragraph{Acknowledgments} We thank M. Caselle for many useful discussions. 
One of us (A. D.) also thanks G. Semenoff for discussions.   
\section*{Appendix}
In this Appendix we display explicitly the coefficients of the
grand-canonical partition
functions for the coverings and the generalized Yang-Mills theory up
to order $q^5$.
\par
Let us consider first $Z^{(\rm cov)}(G,q)$ defined in Eq. (\ref{Zcov}).
Each coefficient $Z^{(\rm cov)}_k(G)$ is the number of inequivalent
unbranched $k$-coverings of a genus $G$ surface without boundaries. Two
coverings are said to be inequivalent if they cannot be obtained from
each other by simple relabeling of the sheets; therefore the total number
of $k$-coverings is $k!Z^{(\rm cov)}_k(G)$.
\par
Naming $\chi=2-2G$ the Euler characteristic of the surface ${\cal M}$,
from Eq. (\ref{Zcov3}) we obtain:
\begin{eqnarray}
Z^{(\rm cov)}_0(G)&=&1~,\nonumber\\
Z^{(\rm cov)}_1(G)&=&1~,\nonumber\\
Z^{(\rm cov)}_2(G)&=&2\cdot 2^{-\chi}~,\nonumber\\
Z^{(\rm cov)}_3(G)&=&2\cdot 6^{-\chi}+3^{-\chi}~,\nonumber\\
Z^{(\rm cov)}_4(G)&=&2\cdot 24^{-\chi}+12^{-\chi}+2\cdot 8^{-\chi}~,\nonumber\\
Z^{(\rm cov)}_5(G)&=&2\cdot 120^{-\chi}+2\cdot 30^{-\chi}+2\cdot 24^{-\chi}+
20^{-\chi}~.
\label{z5}
\end{eqnarray}
The number of inequivalent connected coverings
are given by the coefficients in the power expansion in $q$ of
$\log Z^{(\rm cov)}(G,q)$. Here we give the list up to  $k=5$:
\begin{eqnarray}
F^{(\rm cov)}_1(G)&=&1~,\nonumber\\
F^{(\rm cov)}_2(G)&=&2\cdot 2^{-\chi}-1/2~,\nonumber\\
F^{(\rm cov)}_3(G)&=&2\cdot 6^{-\chi}+3^{-\chi}-2\cdot 2^{-\chi}+1/3~,
\nonumber\\
F^{(\rm cov)}_4(G)&=&2\cdot 24^{-\chi}+12^{-\chi}+2\cdot 8^{-\chi}\nonumber\\
&-&2\cdot 6^{-\chi}-3^{-\chi}+2\cdot 2^{-\chi}-2\cdot 4^{-\chi}-1/4~,
\nonumber\\
F^{(\rm cov)}_5(G)&=&2\cdot 120^{-\chi}+2\cdot 30^{-\chi}+20^{-\chi}
-5\cdot 12^{-\chi}\nonumber\\
&-&2\cdot 8^{-\chi}+4\cdot 4^{-\chi}+3^{-\chi}-
2\cdot 2^{-\chi}+1/5~.
\label{f5}
\end{eqnarray}
By inserting Eqs. (\ref{z5}) and (\ref{f5}) into Eqs. (\ref{freen}) and
(\ref{pf})
we can write the first few terms in the expansion of $Z(G,{\cal A},q)$, namely
the partition functions $Z_{N}({\cal A},G)$ for $N$ up to 5:
\begin{eqnarray}
Z_{0}({\cal A},G)&=&1~,\nonumber\\
Z_{1}({\cal A},G)&=&z({\cal A})~,\nonumber\\
Z_{2}({\cal A},G)&=&\left(2\cdot
2^{-\chi}-\frac{1}{2}\right)z(2{\cal A})+\frac{1}{2}z^{2}({\cal A})~,\nonumber\\
Z_{3}({\cal A},G)&=&\left(2\cdot 6^{-\chi}+3^{-\chi}-2\cdot
2^{-\chi}+\frac{1}{3}\right)z(3{\cal A})
\nonumber\\
&+&\left(2\cdot 2^{-\chi}-\frac{1}{2}\right)z({\cal A})z(2{\cal
A})+\frac{1}{6}z^{3}({\cal A})~,\nonumber \\
Z_{4}({\cal A},G)&=&\Bigl(2\cdot 24^{-\chi}+12^{-\chi}+2\cdot 8^{-\chi}
-2\cdot 6^{-\chi}\Bigr.\nonumber\\
&-&\left.3^{-\chi}+2\cdot 2^{-\chi}-2\cdot 4^{-\chi}-\frac{1}{4}\right)
z(4{\cal A})\nonumber\\
&+&\frac{1}{2}\left(4\cdot 4^{-\chi}-2\cdot 2^{-\chi}+\frac{1}{4}\right)
z^{2}(2{\cal A})\nonumber\\
&+&\left(2\cdot 6^{-\chi}+3^{-\chi}-2\cdot
2^{-\chi}+\frac{1}{3}\right)z({\cal A})z(3{\cal A})\nonumber\\
&+&\frac{1}{2}\left(2\cdot 2^{-\chi}-\frac{1}{2}\right)z(2{\cal A})z^{2}({\cal A})+
\frac{1}{24}z^{4}({\cal A})~,\nonumber\\
Z_{5}({\cal A},G)&=
&\Bigl(2\cdot 120^{-\chi}+2\cdot 30^{-\chi}
+20^{-\chi}
-5\cdot 12^{-\chi}\Bigr.\nonumber\\
&-&\left.2\cdot 8^{-\chi}+4\cdot 4^{-\chi}+3^{-\chi}-
2\cdot 2^{-\chi}+\frac{1}{5}\right)z(5{\cal A})\nonumber\\
&+&\left(2\cdot 24^{-\chi}+12^{-\chi}+2\cdot 8^{-\chi}
-2\cdot 6^{-\chi}\right.\nonumber\\
&-&\left. 3^{-\chi}+2\cdot 2^{-\chi}-2\cdot
4^{-\chi}-\frac{1}{4}\right)z(4{\cal A})z({\cal A})\nonumber\\
&+&\left(4\cdot 12^{-\chi}+6^{-\chi}-4\cdot 4^{-\chi}-\frac{1}{2}\cdot 3^{-\chi}
\right.\nonumber\\
&+& \left.\frac{5}{3}\cdot 2^{-\chi}-\frac{1}{6}\right)z(2{\cal A})z(3{\cal A})
\nonumber\\
&+&\frac{1}{2}\left(2\cdot 6^{-\chi}+3^{-\chi}-2\cdot 2^{-\chi}+\frac{1}{3}
\right)
z^{2}({\cal A})z(3{\cal A})\nonumber\\
&+&\frac{1}{2}\left(4\cdot 4^{-\chi}-2\cdot
2^{-\chi}+\frac{1}{4}\right)z({\cal A})z^{2}(2{\cal A})\nonumber\\
&+&\frac{1}{6}\left(2\cdot 2^{-\chi}-\frac{1}{2}
\right)z^{3}({\cal A})z(2{\cal A})+ \frac{1}{120}z^{5}({\cal A})~,
\end{eqnarray}
where $z({\cal A})$ is a shorthand for $Z_{{\rm U}(1)}({\cal A})$,
namely for the (generalized) U$(1)$ partition function on a surface of area
${\cal A}$:
\begin{equation}
z({\cal A})=\sum_{n}e^{-v(n){\cal A}}~.
\end{equation}
\end{document}